# Gate-induced superconductivity in a monolayer topological insulator

Ebrahim Sajadi[1], Tauno Palomaki[2], Zaiyao Fei[2], Wenjin Zhao[2], Philip Bement[1], Christian Olsen[1], Silvia Luescher[1], Xiaodong Xu[2,3], Joshua A. Folk[1*], and David H. Cobden[2*]

[1]Stewart Blusson Quantum Matter Institute, University of British Columbia, Vancouver BC V6T1Z4, Canada; and Department of Physics and Astronomy, University of British Columbia, Vancouver BC V6T1Z1, Canada

[2]Department of Physics, University of Washington, Seattle WA 98195, USA

[3]Department of Materials Science and Engineering, University of Washington, Seattle, WA

[*]Corresponding authors: cobden@uw.edu, jfolk@physics.ubc.ca

## Abstract:

The layered semimetal $WTe_2$ has recently been found to be a two-dimensional topological insulator (2D TI) when thinned down to a single monolayer, with conducting helical edge channels. We report here that intrinsic superconductivity can be induced in this monolayer 2D TI by mild electrostatic doping, at temperatures below 1 K. The 2D TI-superconductor transition can be easily driven by applying a just a small gate voltage. This discovery offers new possibilities for gate-controlled devices combining superconductivity and topology, and could provide a basis for quantum information schemes based on topological protection.

## Main text:

Many of the most important, and fascinating, phenomena in condensed matter emerge from the quantum mechanics of electrons in a lattice. The periodic potential of the lattice gives rise to Bloch energy bands, and so to the physics of semiconductors that underlies all modern-day electronics. On the more exotic side, electrons in a lattice can pair up to act as bosons and condense into a macroscopic quantum state conducting electricity with zero resistance. More recently, it was realized that Bloch wavefunctions can have a non-trivial topology, incorporating twists analogous to a Möbius strip. This led to the discovery of topological insulators–materials that are electrically insulating in their interior but have conducting boundary modes that result from the topological discontinuity between inside and outside(*1*). The first of these to be studied was the so-called 2D topological insulator (2D TI), in which the one-dimensional helical edge modes (spin locked to momentum) give rise to the quantum spin Hall effect(*2-4*).

Materials that combine non-trivial topology with superconductivity have been the subject of active investigation in recent years. For example, hybrid structures that couple an s-wave superconductor to a 2D TI have also been proposed as platform for Majorana modes(*5*), whose non-abelian exchange properties might be harnessed for qubits(*6*) with coherence times far longer than those built on conventional platforms. There are also topological superconductors, in which vortices or boundaries can host Majorana modes(*7*).

Here we report the remarkable finding that monolayer $WTe_2$, recently shown(*8-13*) to be an intrinsic 2D TI, itself turns superconducting under moderate electrostatic gating. Several other non-topological layered materials superconduct in the monolayer limit, either intrinsically or under heavy doping using ionic liquid gates(*14-22*). However, the present case constitutes the first instance of a phase transition from a 2D topological insulator to a superconductor, which moreover is readily controlled by a gate voltage. The discovery creates new opportunities for gateable superconducting circuitry, and offers the potential to develop topological superconducting devices in a single material, as opposed to the hybrid constructions currently required.

We present data from two monolayer $WTe_2$ devices, M1 and M2, with consistent superconducting characteristics. Each contains a monolayer flake of $WTe_2$ encapsulated along with thin platinum electrical contacts between hexagonal boron nitride (hBN) dielectric layers. Fig. 1A shows an image of M1, which has seven contacts along one edge, together with a side view and a schematic showing the configuration used to measure the linear 4-probe resistance, $R_{xx} = dV/dI$. Top and bottom gates, at voltages $V_t$ and $V_b$ and with areal capacitances $c_t$ and $c_b$ respectively, can be used to induce negative or positive charge in the monolayer $WTe_2$, producing an areal doping density given by $n_e = (c_t V_t + c_b V_b)/e$, where $e$ is the electron charge. Note that we do not interpret this as a carrier density since the insulating state may be of correlated nature (as in, for example, an excitonic insulator); in addition, Hall density measurements are challenging due to the 2D TI edge conduction. More details about gating, contact resistances and capacitances are given in the Supplementary Materials.

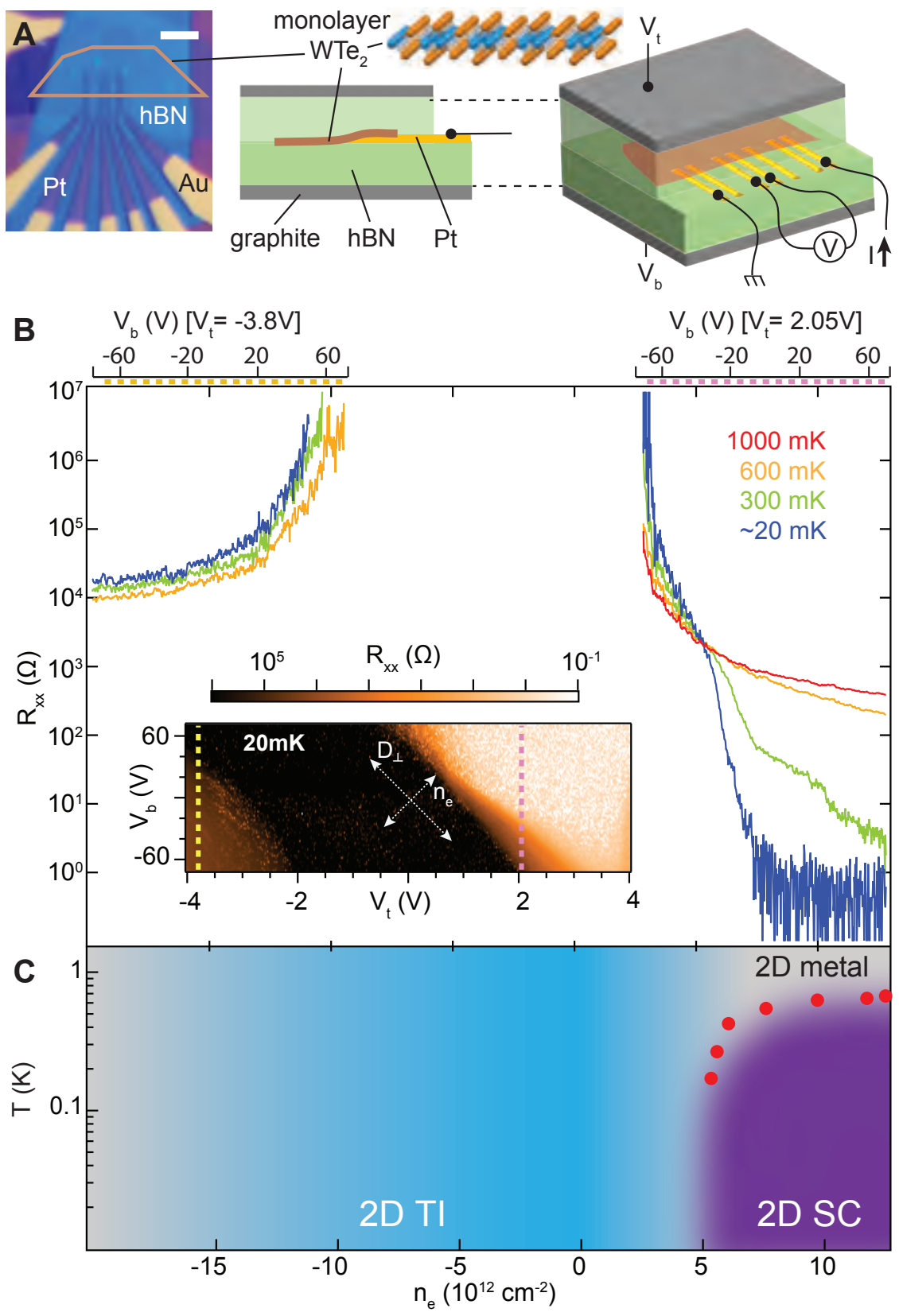

**Figure 1. Characteristics of monolayer $WTe_2$ device M1 at temperatures below 1 K**. (**A**) Optical image (the white scale bar indicates 5 μm) and schematic device structure, showing current, voltage contacts and ground configuration for measuring the 4-probe resistance $R_{xx}$. Inset: atomic structure of monolayer $WTe_2$. (**B**) $R_{xx}$ as a function of electrostatic doping ($n_e$) at a series of temperatures. Inset: variation of $R_{xx}$ at 20 mK with top and bottom gate voltages, $V_t$ and $V_b$, indicating the axes corresponding to doping $n_e$ and transverse displacement field $D_\perp$. $R_{xx}$ depends primarily on $n_e$ and only weakly on $D_\perp$. The measurements in the main panel for $n_e > 0$ and $n_e < 0$ were made separately, sweeping $V_b$ along the two colored dashed lines to avoid contact effects. (**C**) Phase diagram constructed from measurements in this paper.

Figure 1B illustrates electrostatic tuning M1 from p-doped conducting behaviour at negative gate voltage, through the 2D TI state(*9, 13*), to an n-doped highly conducting state at positive gate voltage. $R_{xx}$ in the 2D TI state, around $n_e = 0$, is more than $10^7$ Ω due to a meV-scale gap that blocks edge conduction below 1K (see later in the text and Supplementary Materials). For $n_e$ above $n_{crit} \approx +5 \times 10^{12}$cm$^{-2}$, however, the resistance drops dramatically when the sample is cooled, reaching the noise floor of the experiment (around 0.3 Ω) for $n_e > +7 \times 10^{12}$ cm$^{-2}$ at 20 mK indicating the appearance of superconductivity. Figure 1C is a phase diagram constructed from these and similar measurements discussed below. The emergence of a superconducting phase in direct proximity to a 2D TI phase, and at a doping level achievable with a single electrostatic gate, is the primary result of this paper.

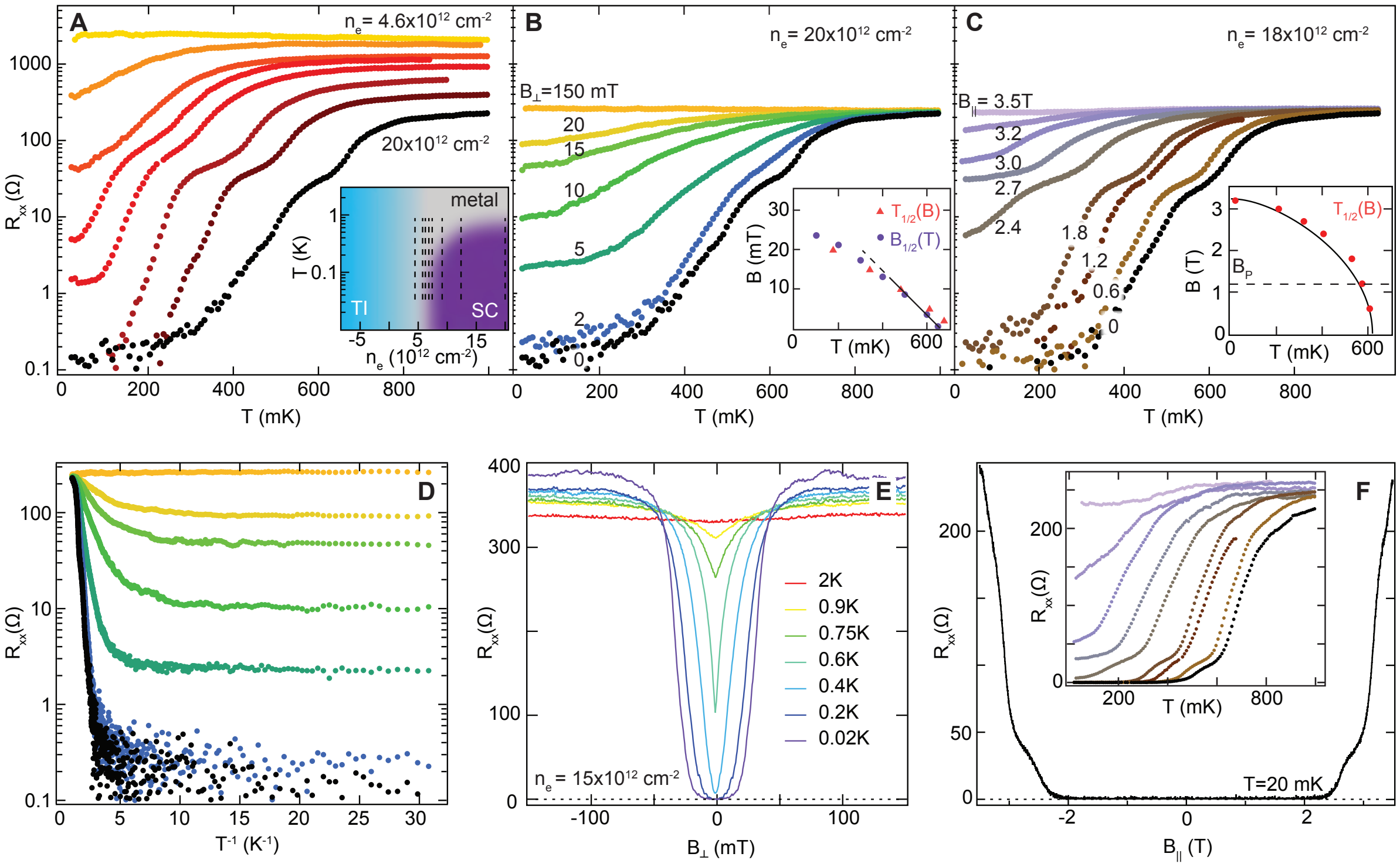

**Figure 2**. **Resistance characterization of device M1 in the superconducting regime**. (**A**) $R_{xx}$ on log scale vs temperature $T$ at a series of positive-gate doping levels $n_e$ [20, 12, 8.5, 6.7, 6.1, 5.6, 5, 4.6 × $10^{12}$ cm$^{-2}$] showing a drop of several orders of magnitude at low $T$ for larger $n_e$. Inset: location of sweeps on the phase diagram. (**B**) Effect of perpendicular magnetic field, $B_\perp$ on resistance at the highest $n_e$ in panel **A**. (Demagnetization effects are neglected in light of the finite resistivity of the sample.) Inset: characteristic temperatures $T_{1/2}$ obtained from these temperature sweeps, as well as characteristic fields $B_{1/2}$ measured from field sweeps under similar conditions. (**C**) Same for in-plane magnetic field, $B_\parallel$. ($B_\parallel = 0$ data are for $n_e = 19 \times 10^{12}$ cm$^{-2}$). Inset shows reduction of $T_{1/2}$ with $B_\parallel$, fit to the expected form for materials with strong spin-orbit scattering (solid line). The Pauli limit $B_P$, assuming $g = 2$, is indicated by the dashed line. (**D**) Data from panel **B** replotted to highlight the saturation of $R_{xx}$ at low $T$. (**E**) Sweeps of $B_\perp$ showing rise of resistance beginning at very low field. (**F**) Sweep of $B_\parallel$ showing sharper onset of resistance compared to (**E**). Inset: data from (**C**) on a linear scale.

The transition from an insulating to a metallic/superconducting $T$ dependence—the crossing of $R_{xx}$ lines in Fig. 1B—occurs at 2.4 kΩ. This corresponds to a square resistivity $\rho \approx 20$ kΩ, with a substantial uncertainty because the precise distribution of current in the device is not known (see Supplementary Materials). The evolution of the $T$ dependence with $n_e$ is illustrated in Fig. 2A. For all densities shown, the collapse of $R_{xx}$ with temperature is gradual, as expected for materials where the normal state 2D conductivity is not much greater than $e^2/h$. We define a characteristic temperature, $T_{1/2}$, at which $R_{xx}$ falls to half its 1 K value. Although this specific definition is somewhat arbitrary, it is typical in the literature(*15, 21, 22*), and does not significantly affect any of our conclusions (see Supplementary Materials). Measured values of $T_{1/2}$ are plotted on the phase diagram in Fig. 1C to indicate boundaries of superconducting behavior.

The superconductivity is suppressed by perpendicular ($B_\perp$) or in-plane ($B_\parallel$) magnetic field, as illustrated in Figs. 2B and 2C respectively. For perpendicular field, orbital effects are expected to dominate(*23-25*). The dependence of $T_{1/2}$ on $B_\perp$ (Fig. 2B inset) in the low-field limit is consistent with the linear $B_{c2}^{\perp}(T)$ expected from Ginzburg-Landau theory. The characteristic perpendicular field in the low temperature limit is $B_{1/2}^{\perp}(T \to 0) \approx 25$ mT, based on the measurements in Fig. 2B (inset). Estimates for the superconducting coherence length can be obtained either from the slope of $B_{1/2}^{\perp}(T)$ near $T_{1/2}$ or from $B_{1/2}^{\perp}(T \to 0)$, yielding $\xi_{meas} = 100 \pm 30$ nm in both cases (see Supplementary Materials).

The fact that $\xi_{meas}$ is significantly larger than the estimated mean free path $\lambda = h/\left(e^2\rho\sqrt{g_s g_v \pi n_e}\right) \approx 8$ nm suggests that the dirty limit $\lambda \ll \xi$ applies. To calculate $\lambda$ we use spin and valley degeneracies $g_s = g_v = 2$ , and density and normal-state resistivity reflecting the conditions for Fig. 2B, $n_e = 19 \times 10^{12}$ cm$^{-2}$ and $\rho \approx 2$ kΩ respectively. The coherence length expected in the dirty limit is $\xi \approx \sqrt{\hbar D/\Delta_0}$, for zero-temperature gap $\Delta_0 = 1.76\, k_B T_c$ and diffusion constant $D$. Indeed, using $T_{1/2} = 700$ mK for $T_c$, and $D = 2\pi\hbar^2/g_s g_v m^* e^2 \rho \approx 12$ cm$^2$s$^{-1}$ (from the Einstein relation) with effective mass(*26*) $m^* = 0.3 m_e$ gives $\xi \approx 90$ nm, consistent with $\xi_{meas}$.

For in-plane field, the atomic thinness of the monolayer makes orbital effects small. In the absence of spin scattering, superconductivity is then suppressed when the energy associated with Pauli paramagnetism in the normal state overcomes the superconducting condensation energy. This is referred to as the Pauli (Chandrasekar-Clogston) limit(*27*), and gives a critical field $B_P = 1.76 k_B T_c / g^{1/2} \mu_B$. Assuming an electron g-factor of $g = 2$ and taking $T_c = 700$ mK gives $B_P \approx 1.3$ T. However, the data in Figs. 2C and 2F indicate superconductivity persisting to $B_{1/2}^{\parallel} = 3$ T.

Similar examples of $B_{1/2}^{\parallel}$ exceeding $B_P$ have recently been reported in other monolayer dichalcogenides, $MoS_2$ and $NbSe_2$, but the Ising superconductivity mechanism(*15, 21*) invoked in those works cannot explain an enhancement of $B_{1/2}^{\parallel}$ here because $WTe_2$ lacks the required in-plane mirror symmetry. One possible explanation in this case is a high spin-orbit scattering rate $\tau_{so}^{-1}$. Fitting the predicted form for $T_c$ in a parallel field(*28*) to the data in the inset to Fig. 2C gives $\tau_{so}^{-1} \approx 2$ ps$^{-1}$ (see Supplementary Materials). Another possibility is that the Pauli limit is not actually exceeded, but that the effective g-factor in $WTe_2$ is smaller than 2 due to the strong spin-orbit coupling.

The data in Fig. 2 display several other features worthy of mention. First, at intermediate magnetic fields the resistance approaches a $T$-independent level as $T \to 0$ that is orders of magnitude below the normal-state resistance. The data from Fig. 2B are replotted vs $1/T$ in Fig. 2D to highlight the behavior below 100 mK. Similar behavior is seen at $B = 0$ (Fig. 2A) for intermediate $n_e$, adding to the growing body of evidence that this is a robust phenomenon occurring in thin films close to superconductivity(*29*). Second, even at the lowest temperature, $R_{xx}$ rises smoothly from zero as a function of $B_\perp$ (Fig. 2E) whereas the onset of measurable resistance as a function of $B_\parallel$ is relatively sudden, occuring above 2 T (Fig. 2F). Third, an intermediate plateau is visible in the $R_{xx} - T$ data at $B = 0$ over a wide range of $n_e$ (Fig. 2A). It is extremely sensitive to $B_\perp$, almost disappearing at only 2 mT (Fig. 2B), whereas it survives in $B_\parallel$ to above 2 T (Fig. 2C and inset to Fig. 2F). A similar feature has been reported in some other quasi-2D superconductors(*30-32*), but its nature, and the role of disorder, remain unresolved.

The high tunability of this new 2D superconducting system invites comparison with theoretical predictions for critical behavior close to a quantum phase transition. Figure 3 shows how $R_{xx}$ depends on doping at a series of temperatures, along the dashed lines in the inset phase diagram.

The $T$ dependence changes sign at $n_{crit} \approx 5 \times 10^{12}$ cm$^{-2}$. In the inset we show an attempt to collapse the data onto a single function of $|1 - n_e/n_{crit}|\, T^{-\alpha}$. The procedure is somewhat hindered by the fluctuations, which can be seen to be largely reproducible. The best-fit critical exponent $\alpha = 0.8$ is similar to that reported for some insulator-superconductor transitions in thin films(*33*), although we note that the anomalous metallic behavior near $n_{crit}$ is not consistent with such a scaling.

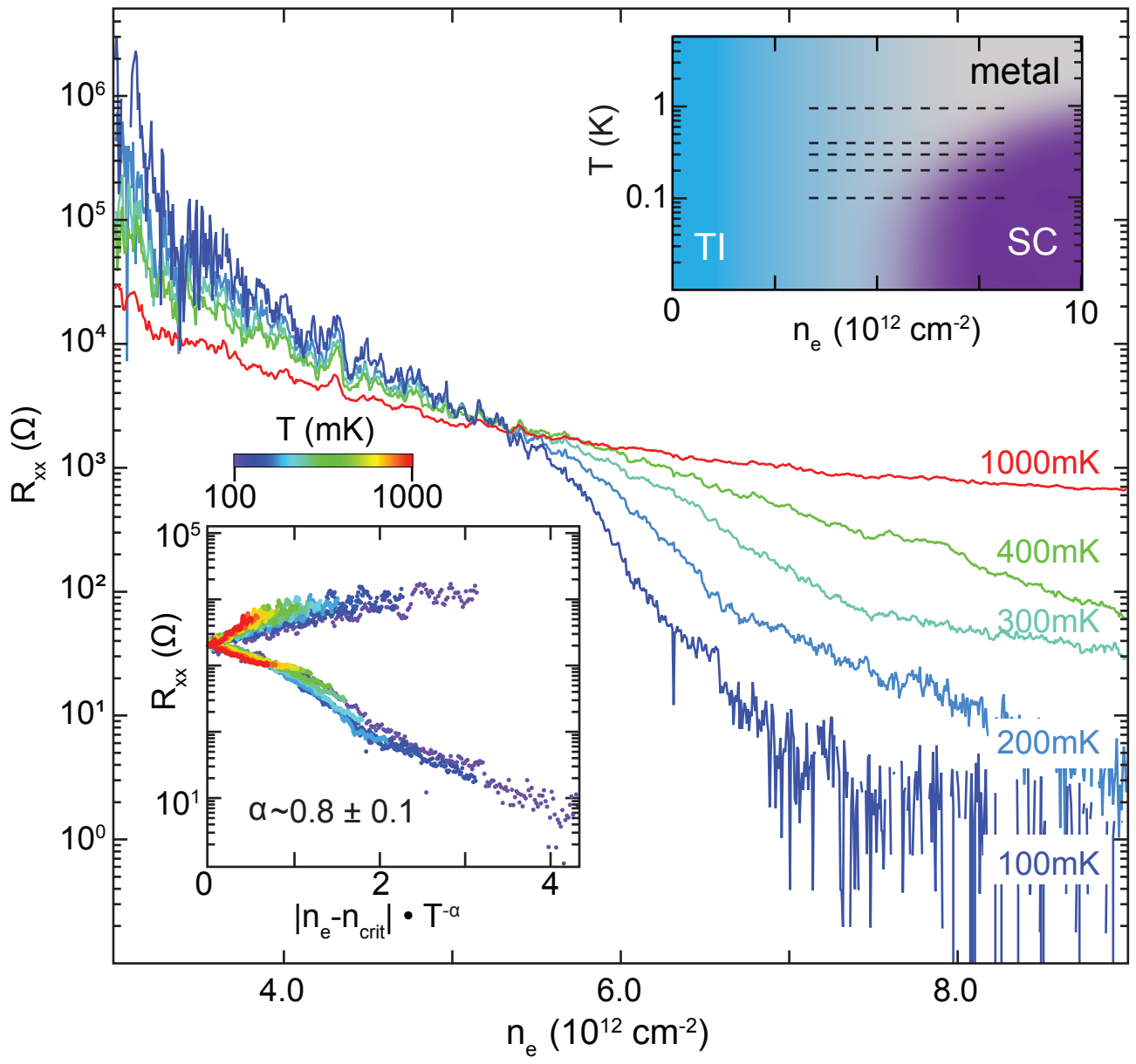

**Figure 3. Scaling analysis of the transition.** Main panel: Multiple $R_{xx}$ vs doping traces, taken at different temperatures, cross at a critical doping level $n_{crit} \approx 5 \times 10^{12}$ cm$^{-2}$. Upper inset: dashed lines locate these sweeps on the phase diagram. Lower inset: same data presented on a scaling plot, taking critical exponent $\alpha = 0.8$.

Superconductivity induced by simple electrostatic gating in a monolayer of material that is not normally superconducting is intriguing, but perhaps even more interesting is that the ungated state is a 2D TI. This prompts the question of whether the helical edge channels remain when the superconductivity appears, and if so, how strongly they couple to it. In principle, $R_{xx}$ includes contributions from both edges and bulk. However, since in device M1 the edge conduction freezes out below 1 K, in order to investigate the combination of edge channels and superconductivity we turn to another device, M2, in which edge conduction persists to lower temperatures (Supplementary Materials).

Figure 4 shows measurements of the conductance, $G$, between adjacent contacts in M2 as a function of gate doping. The figure includes schematics indicating the inferred state of the edge (red for conducting), as well as the bulk state (colored to match the phase diagram). Consider first the black trace, taken at 200 mK and $B_\perp = 0$. At low $n_e$ the bulk is insulating and edge conduction dominates, albeit with large mesoscopic fluctuations. For $n_e > 2 \times 10^{12}$ cm$^{-2}$, $G$ increases as bulk conduction begins, then once $n_e$ exceeds $n_{crit}$ it increases faster as superconductivity appears, before leveling out at around 200 μS due to contact resistance. This interpretation is supported by warming to 1 K (red dotted trace), which destroys the superconductivity and so reduces $G$ for $n_e > n_{crit}$, but enhances the edge conduction at low $n_e$ towards the ideal value of $e^2/h$ =39 μS. (We note that this $T$ dependence of the edge is associated with a gap of around 100 μeV, visible in the inset map of differential conductance vs bias and doping). A perpendicular field $B_\perp$ of 50 mT (green trace) also destroys the superconductivity, causing the conductance to fall for $n_e > n_{crit}$ but barely affecting it at lower $n_e$. High magnetic fields have been shown(*9*) to suppress edge conduction in the 2D TI state due to breaking of time reversal symmetry. This effect can be clearly seen in the $B_{\perp} = 1$ T data (orange trace in Fig. 4) as $G$ falls to zero at low $n_e$. Importantly, $G$ falls by a similar amount at higher $n_e$, implying that the edge conduction supplies a parallel contribution, and thus that the helical edge states persist when $n_e > n_{crit}$ and at temperatures below $T_c$.

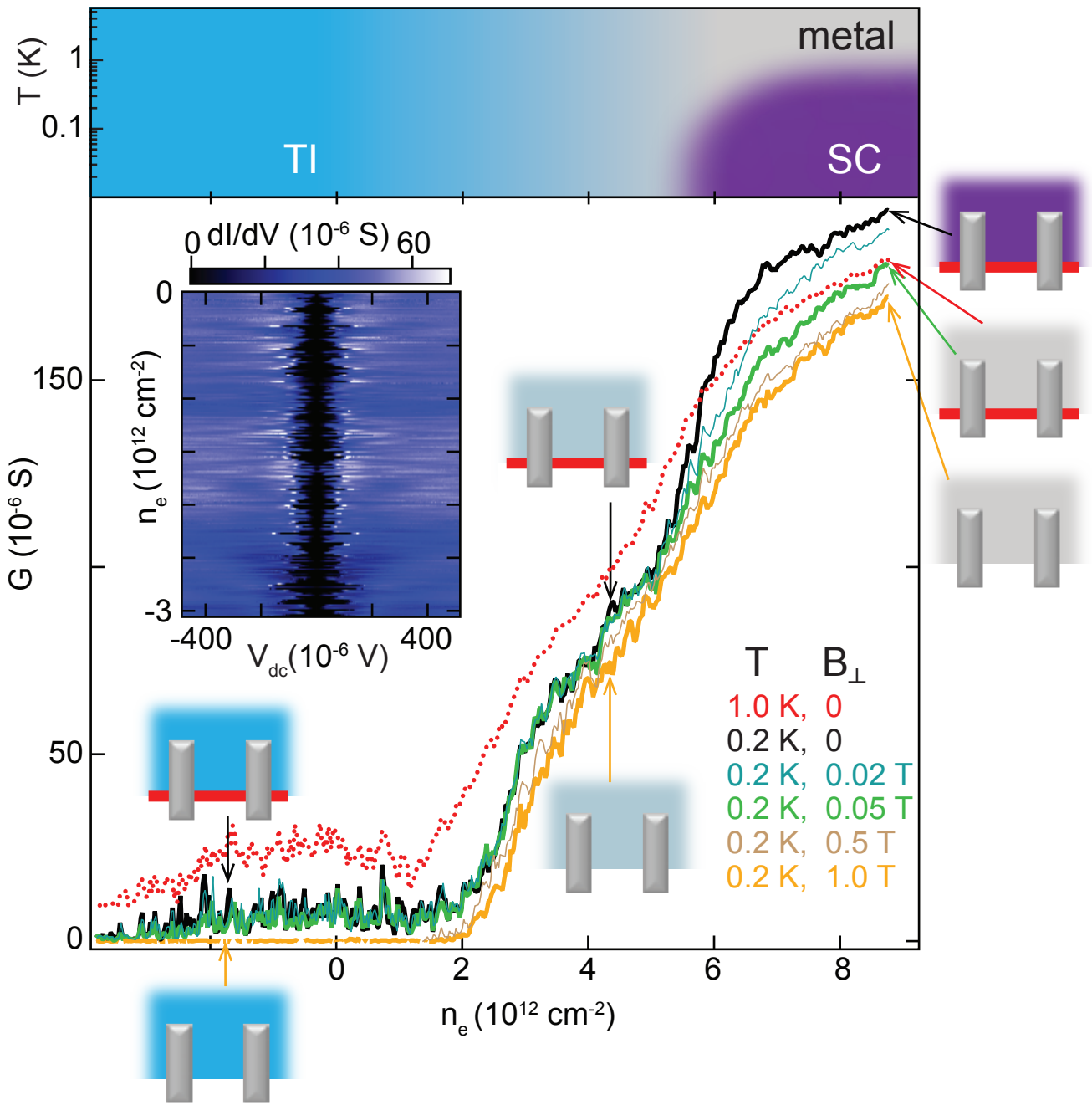

**Figure 4. Evidence for the presence of both edge conduction and superconductivity in device M2**. The main panel shows the linear conductance between two adjacent contacts vs gate doping at the temperatures and perpendicular magnetic fields noted. Schematics indicate the state of edge and bulk conduction at different points, the bulk being colored to match the phase diagram reproduced above. Superconductivity occurs for $n_e > 5 \times 10^{12}$ cm$^{-2}$ at $B = 0$; edge conduction dominates for $n_e < 2 \times 10^{12}$ cm$^{-2}$ but appears to be present at all $n_e$. Inset: color-scale plot of differential conductance vs dc voltage bias and doping level, revealing a gap of around 100 μeV that fluctuates rapidly as a function of doping level.

This discovery raises compelling questions for future investigation. It is likely that the helical edge modes persist when the superconductivity is restored by reducing the magnetic field to zero. Other techniques, such as scanning tunneling microscopy or contacts discriminating edge from bulk, may be needed to probe the edges distinct from the bulk. The measurements presented here cannot determine the degree or nature of the coupling between superconductivity and edge conduction. One key question moving forward is whether the edge states also develop a superconducting gap, in which case they could host Majorana zero modes(*5*).

Another question concerns the nature of the superconducting order. It is striking that $n_{crit}$ corresponds to only ~0.5% of an electron per W atom, which is at least ten times lower than the doping level needed to observe superconductivity in other transition metal dichalcogenide monolayers(*18*). Many-layer $WTe_2$ is semimetallic(*34-37*) under ambient conditions, with near-perfect compensation of electrons and holes, but becomes superconducting as the ratio of electrons to holes increases at high pressure(*38*). Some related materials, such as $TiSe_2$, are known to switch from charge-density-wave to superconducting states at quite low doping(*39*) or under pressure(*40*). We therefore speculate that doping tips the balance in monolayer $WTe_2$ in favor of superconductivity, away from a competing insulating electronic ordering. Finally, given the topological band structure and likely strong correlations in this material, it is possible that the pairing is unconventional and perhaps topologically nontrivial.

## Acknowledgements

We thank Oded Agam, Anton Andreev and Boris Spivak for discussions, and Jiaqiang Yan for the $WTe_2$ crystals. TP, ZF, WZ, XX, and DC supported by the U.S. Department of Energy, Office of Basic Energy Sciences, Division of Materials Sciences and Engineering, Awards DE-SC0002197 (DHC) and DE-SC0018171 (XX); AFOSR FA9550-14-1-0277; NSF EFRI 2DARE 1433496; and NSF MRSEC 1719797. ES, PB, CO, SL and JF supported by the Canada Foundation for Innovation; the National Science and Engineering Research Council; CIFAR; and SBQMI.

## References

1. M. Z. Hasan, C. L. Kane, *Colloquium*: Topological insulators. *Reviews of Modern Physics* **82**, 3045 (2010).
2. B. A. Bernevig, T. L. Hughes, S.-C. Zhang, Quantum Spin Hall Effect and Topological Phase Transition in HgTe Quantum Wells. *Science* **314**, 1757-1761 (2006).
3. C. Kane, E. Mele, $Z_2$ topological order and the quantum spin Hall effect. *Physical Review Letters* **95**, (2005).
4. M. König *et al.*, Quantum Spin Hall Insulator State in HgTe Quantum Wells. *Science* **318**, 766-770 (2007).
5. L. Fu, C. L. Kane, Superconducting Proximity Effect and Majorana Fermions at the Surface of a Topological Insulator. *Physical Review Letters* **100**, 096407 (2008).
6. S. Sarma, M. Freedman, C. Nayak, Majorana zero modes and topological quantum computation. *Npj Quantum Information* **1**, (2015).
7. A. Masatoshi Sato and Yoichi, Topological superconductors: a review. *Reports on Progress in Physics* **80**, 076501 (2017).
8. X. Qian, J. Liu, L. Fu, J. Li, Quantum spin Hall effect in two-dimensional transition metal dichalcogenides. *Science* **346**, 1344-1347 (2014).
9. Z. Fei *et al.*, Edge conduction in monolayer $WTe_2$. *Nat Phys* **13**, 677-682 (2017).
10. S. Tang *et al.*, Quantum spin Hall state in monolayer 1T'-$WTe_2$. *Nat Phys* **13**, 683-687 (2017).
11. L. Peng *et al.*, Observation of topological states residing at step edges of $WTe_2$. *Nature Communications* **8**, 659 (2017).
12. Z.-Y. Jia *et al.*, Direct visualization of a two-dimensional topological insulator in the single-layer 1T $WTe_2$. *Physical Review B* **96**, 041108 (2017).
13. S. Wu *et al.*, Observation of the quantum spin Hall effect up to 100 kelvin in a monolayer crystal. *Science* **359**, 76-79 (2018).
14. A. W. Tsen *et al.*, Nature of the quantum metal in a two-dimensional crystalline superconductor. **12**, 208 (2015).
15. X. Xi *et al.*, Ising pairing in superconducting $NbSe_2$ atomic layers. *Nature Physics* **12**, 139-143 (2016).
16. Q. Y. Wang *et al.*, Interface-Induced High-Temperature Superconductivity in Single Unit-Cell FeSe Films on $SrTiO_3$. *Chinese Physics Letters* **29**, 4 (2012).
17. L. J. Li *et al.*, Controlling many-body states by the electric-field effect in a two-dimensional material. *Nature* **529**, 185-189 (2016).
18. Y. Fu *et al.*, Gated tuned superconductivity and phonon softening in monolayer and bilayer $MoS_2$. *npj Quantum Materials* **2**, 52 (2017).
19. Y. Saito, T. Nojima, Y. Iwasa, Gate-induced superconductivity in two-dimensional atomic crystals. *Superconductor Science and Technology* **29**, 093001 (2016).

20. K. Ueno *et al.*, Electric-field-induced superconductivity in an insulator. *Nature Materials* **7**, 855-858 (2008).
21. J. M. Lu *et al.*, Evidence for two-dimensional Ising superconductivity in gated $MoS_2$. *Science* **350**, 1353 (2015).
22. D. Costanzo, S. Jo, H. Berger, A. F. Morpurgo, Gate-induced superconductivity in atomically thin $MoS_2$ crystals. *Nature Nanotechnology* **11**, 339 (2016).
23. R. A. Klemm, Pristine and intercalated transition metal dichalcogenide superconductors. *Physica C: Superconductivity and its Applications* **514**, 86-94 (2015).
24. M. Tinkham, *Introduction to superconductivity*. International series in pure and applied physics (McGraw Hill, New York, ed. 2nd, 1996), pp. xxi, 454 pages.
25. Y. Saito, T. Nojima, Y. Iwasa, Highly crystalline 2D superconductors. *Nature Reviews Materials* **2**, 16094 (2016).
26. H. Y. Lv *et al.*, Perfect charge compensation in $WTe_2$ for the extraordinary magnetoresistance: From bulk to monolayer. *EPL* **110**, 37004 (2015).
27. A. M. Clogston, Upper limit for critical field in hard superconductors. *Physical Review Letters* **9**, 266-& (1962).
28. R. A. Klemm, A. Luther, M. R. Beasley, Theory of the upper critical field in layered superconductors. *Physical Review B* **12**, 877-891 (1975).
29. A. Kapitulnik, S. A. Kivelson, B. Spivak, Anomalous metals - failed superconductors. *ArXiv e-prints* **1712**, arXiv:1712.07215 (2017).
30. Y. Cheng, M. B. Stearns, Superconductivity of Nb/Cr multilayers. *Journal of Applied Physics* **67**, 5038-5040 (1990).
31. L. Z. Deng *et al.*, Evidence for defect-induced superconductivity up to 49 K in $(Ca_{1-x}R_x)Fe_2As_2$. *Physical Review B* **93**, 054513 (2016).
32. S.-G. Jung *et al.*, Effects of magnetic impurities on upper critical fields in the high- $T_c$ superconductor La-doped $CaFe_2As_2$. *Superconductor Science and Technology* **30**, 085009 (2017).
33. A. M. Goldman, Superconductor-insulator transitions. *International Journal of Modern Physics B* **24**, 4081-4101 (2010).
34. A. A. Soluyanov *et al.*, Type-II Weyl semimetals. *Nature* **527**, 495-498 (2015).
35. M. N. Ali *et al.*, Large, non-saturating magnetoresistance in $WTe_2$. *Nature* **514**, 205-208 (2014).
36. D. MacNeill *et al.*, Control of spin-orbit torques through crystal symmetry in $WTe_2$/ferromagnet bilayers. *Nature Physics* **13**, 300- (2017).
37. V. Fatemi *et al.*, Magnetoresistance and quantum oscillations of an electrostatically tuned semimetal-to-metal transition in ultrathin $WTe_2$. *Physical Review B* **95**, 041410(R) (2017).
38. D. F. Kang *et al.*, Superconductivity emerging from a suppressed large magnetoresistant state in tungsten ditelluride. *Nature Communications* **6**, 6 (2015).
39. E. Morosan *et al.*, Superconductivity in $Cu_xTiSe_2$. **2**, 544 (2006).
40. A. F. Kusmartseva, B. Sipos, H. Berger, L. Forró, E. Tutiš, Pressure Induced Superconductivity in Pristine 1T-$TiSe_2$. *Physical Review Letters* **103**, 236401 (2009).

# Supplementary Materials: Gate-induced superconductivity in a monolayer topological insulator

Ebrahim Sajadi, Tauno Palomaki, Zaiyao Fei, Wenjin Zhao, Philip Bement, Christian Olsen, Silvia Luescher, Xiaodong Xu, Joshua Folk, and David H. Cobden

**This PDF file includes:**

## S1. Device details

Measurements on two devices are presented in the main text, M1 and M2. The making of device M1 began with exfoliated few-layer hBN (lower hBN) on thermally grown $SiO_2$ on a highly doped Si substrate. Seven metal contacts in a row were defined on the lower hBN using electron beam lithography (EBL) and metallized (~7 nm Pt) in an e-beam evaporator followed by acetone lift-off and annealing at 200 °C. A second exfoliated few-layer hBN (upper hBN) flake was picked up using a polymer-based dry transfer technique(*1*). Flux-grown $WTe_2$ crystals (from Jiaqiang Yan, Oak Ridge National Lab) were exfoliated in the glovebox ($O_2$ and $H_2O$ concentrations < 0.5 ppm) and monolayer pieces were optically identified. After identification, the monolayer $WTe_2$ was picked up on the upper hBN and transferred to the Pt/hBN/$SiO_2$/Si stack. Only after fully encapsulating the $WTe_2$ was the device removed from the glovebox. After dissolving the polymer a few-layer graphene (FLG) flake was added as the top gate. Finally, contact pads consisting of ~70/7 nm Au/V were added.

Device M2 began with few-layer hBN (lower hBN) covering a FLG bottom gate on thermally grown $SiO_2$ prepared using the same dry transfer technique and then annealed at 400 C. As in M1, contacts were patterned on the lower hBN and ~7 nm Pt was evaporated, but now in a Hall bar configuration, and the upper hBN flake was prepared. Monolayer $WTe_2$ was picked up with upper hBN and put down on the Pt contacts in the glove box. Finally, for this device, a Hall bar-shaped top gate was patterned with EBL together with the contact pads, using ~7/70 nm evaporated Au/V, followed by lift-off in acetone. The full device structure is shown in Figs. S1C and S1D.

The thickness of each hBN flake was determined using an AFM (Bruker Dimension Edge), and is listed in Table 1 along with the calculated geometric areal capacitance for the top and bottom gates for both devices, given by $c = \frac{\varepsilon_0 \varepsilon_r}{d}$, where we take $\varepsilon_r = 4$ for hBN and 3.9 for $SiO_2$. The total capacitances $c_t$ and $c_b$ defined in the main text are close to the geometric values as long as the applied gate voltages are substantially larger than the gap in the spectrum at zero doping. This gap is known(*2*) to be no more than about 50 meV.

| Device Label | upper hBN (nm) | lower hBN (nm) | $SiO_2$ (nm) | $c_t$ ($nF/cm^2$) | $c_b$ ($nF/cm^2$) |
|---|---|---|---|---|---|
| M1 | 5.8 | 18 | 285 | 611 | 11.4 |
| M2 | 8 | 29 | - | 443 | 122 |

Table 1: Thicknesses of hBN for both M1 and M2 as measured by AFM, and capacitances per unit area for bottom and top gates for both devices.

Gating in M1 was uniform over the entire $WTe_2$ flake, except due to screening of the bottom gate by the Pt contacts. For M2, on the other hand, the combination of a patterned top gate and a bottom gate that covered only part of the $WTe_2$ flake made gate-defined conducting regions more complicated. The two contacts used for the conductance measurements in Fig. 4 (main text) are indicated in Fig. S1F. For the data in Fig. 4 in the main text, the graphene bottom gate was used to induce n-type conductivity between the contacts, while the top gate (shaped like a Hall bar) was fixed at -1.5 V to suppress conduction in the center of the flake. As a result, conductance between the two active contacts was dominated by the pink region of $WTe_2$ highlighted in Fig. S1F, in parallel with edge conduction on the diagonal edge that connects the two contacts (red dashed line in Fig. S1F).

There are two places in the main text where a sheet resistivity is estimated based on the measured $R_{xx}$ in M1, a calculation that requires an aspect ratio and/or knowledge of how the current flows through the device. The numbers presented in the main text assume that the current flows primarily in the rectangle between source and drain current contacts indicated in Fig. S1B, with minimal spreading. Then, the conversion from resistance to resistivity requires only the aspect ratio of the $WTe_2$ between voltage probes, which is estimated to be 10 (width/length) from the AFM image in Fig. S1B.

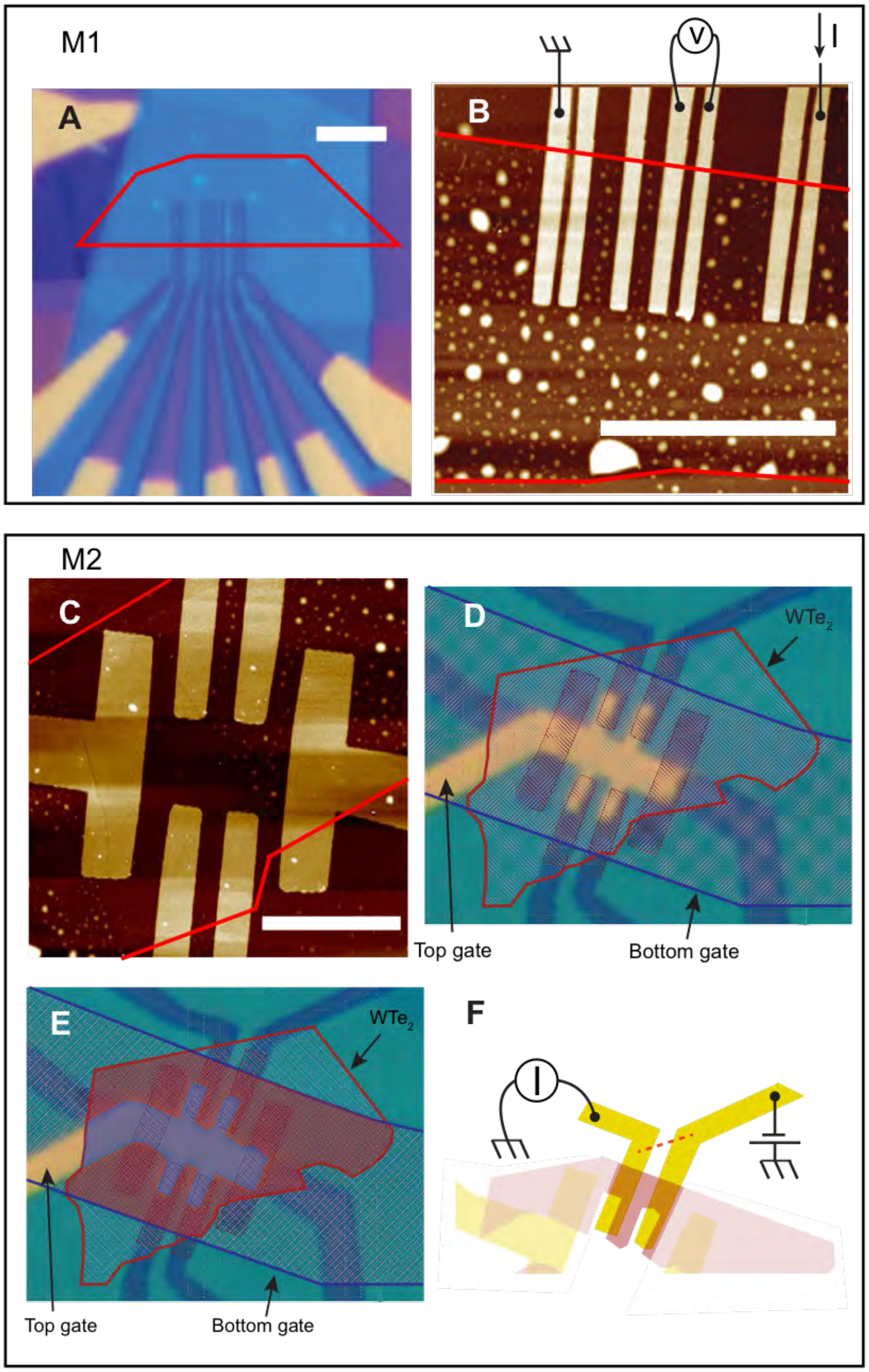

Fig S1. Optical and AFM images of devices M1 and M2. The boundaries of $WTe_2$ are indicated by red lines. **A.** Optical image of device M1. **B.** AFM image of device M1. **C.** AFM image of device M2. **D**. Optical image of device M2. **E.** Optical image of device M2. The region of $WTe_2$ with higher conductivity due to the combined effects of top and bottom gates is shown in light red (see text). **F.** Schematic of the region of interest in M2, showing the contacts used for measurements in Fig. 4 (main text) and highlighting in red the region of $WTe_2$ that dominates the conductance between those two contacts. Note that the physical edge of the flake that connects those two contacts, shown by a red dashed line, is not above the bottom gate. However, based on the detailed properties seen in the data in Fig. 4 we deduce that the conduction at low $n_e$ is dominated by one or more internal cracks, not visible in the images above, that exist in the region above the bottom gate. All scale bars indicate 5μm.

S2. Contact resistance of device M1

The $WTe_2$ flakes in both devices lay on top of the Pt contacts, so top and bottom gates affected the bulk carrier density and contact resistance differently (see schematic in Fig. 1 in the main text). The Pt contacts screen the electric field from the bottom gate, so the bottom gate voltage ($V_b$) only affects the $WTe_2$ carrier density away from the contacts. The top gate voltage, on the other hand, affects the $WTe_2$ everywhere and therefore has a stronger effect on the contact resistance. In order to keep the contact resistance low while changing the carrier density, the voltage applied to the top gate ($V_t$) was fixed at a large positive value for n-type transport—or a large negative value for p-type transport—while varying the voltage applied to the bottom gate ($V_b$).

Figure S2 illustrates the effects of both $V_t$ and $V_b$ on contact resistances for device M1. This figure shows a contact resistance measurement for one particular pair of contacts, as an example for a behavior that was generic for all contacts. In this case, we measured the contact resistance of the contacts, labeled I and G respectively, to which the current bias and ground are connected in the schematic. The device was current biased and two voltage drops were measured simultaneously: $V_1$ records, effectively, the voltage drop within the $WTe_2$ film between contacts I and $G$, whereas $V_2$ includes the voltage drop across the contacts. Thus the contact resistance per contact can be estimated from $R_c = (V_2/I - V_1/I)/2$. In Fig. S2 this quantity is plotted versus $n_e$ for fixed $V_b$, although the data was taken by sweeping $V_t$ for fixed $V_b$.

As seen in the figure, and described in more detail in the caption, maintaining lower contact resistances at moderate to low $n_e$ mandates working at more positive top gate voltages, that is, at more negative back gate voltages for a given density, for the n-doped regime shown here.

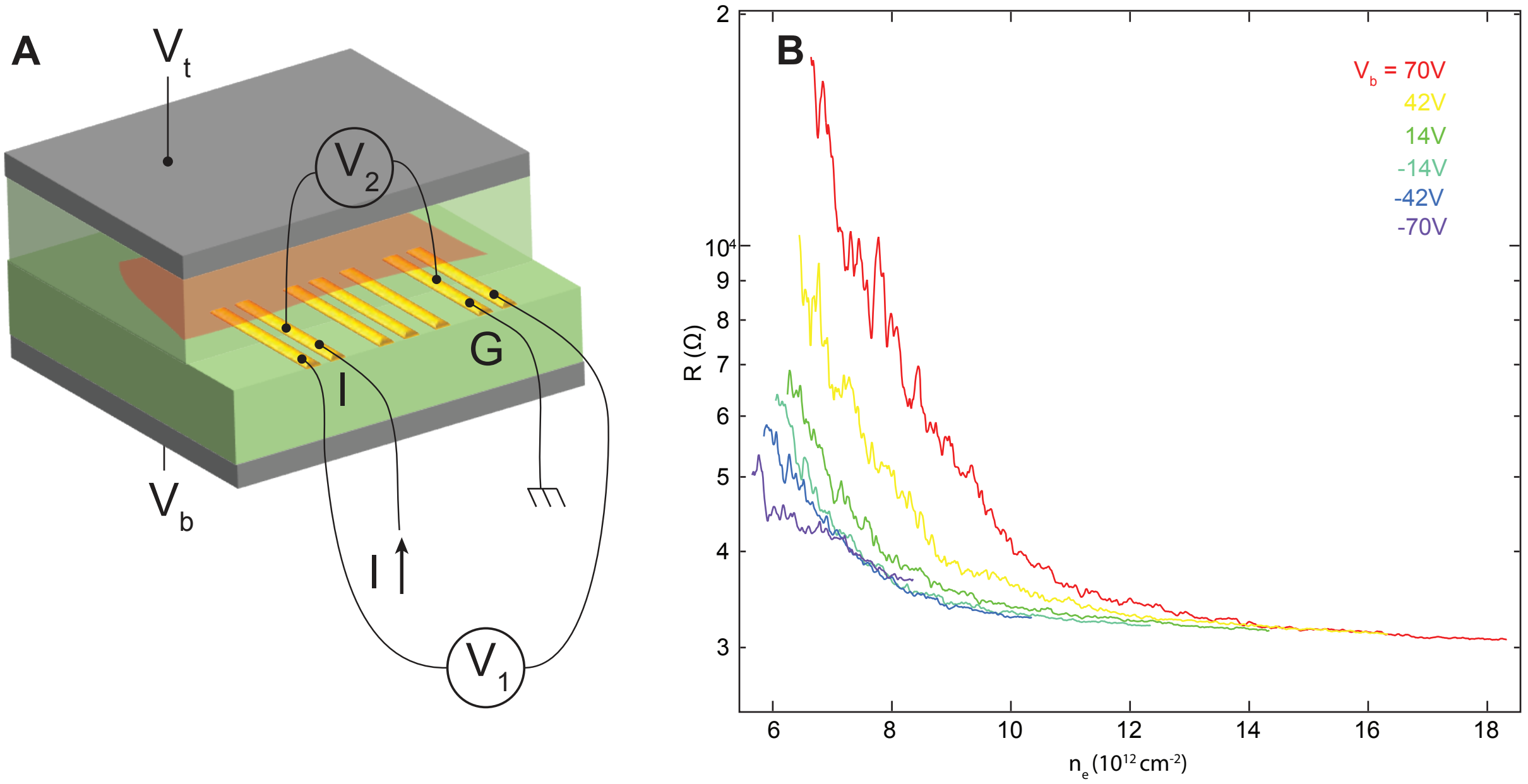

Fig S2. **A.** Contact resistance measurement for a particular pair of contacts (in this case, those to which the current bias and ground are connected, labelled $I$ and $G$). **B.** Although the data is plotted vs $n_e$, the curves represent sweeps of $V_t$ for various fixed $V_b$. When the density is high ($n_e > 12 \times 10^{12}$ cm$^{-2}$), contact resistance is low independent of relative top- and bottom- gate voltages. For lower densities ($n_e < 8 \times 10^{12}$ cm$^{-2}$), however, the contact resistance is much lower for strongly negative bottom gate, that is, where the top gate is very positive. Conversely, less positive top gate voltages (corresponding to more positive bottom gate voltages) give very high contact resistances at lower density.

S3. Plateau in R vs T data

The data in the main text (especially Fig. 2) exhibit a weak intermediate plateau-like feature in resistance, as superconductivity is suppressed either by increasing temperature or in-plane field. In this section, we present data (Fig. S3) for the analogous plateau in the neighboring set of voltage contacts (using the same current source and ground). The similarity of this feature, compared to Fig. 2A, indicates that the plateau is not due to a single localized defect. However, other contact pairs on the same sample showed either no plateau, a plateau at a different level, or multiple plateaus.

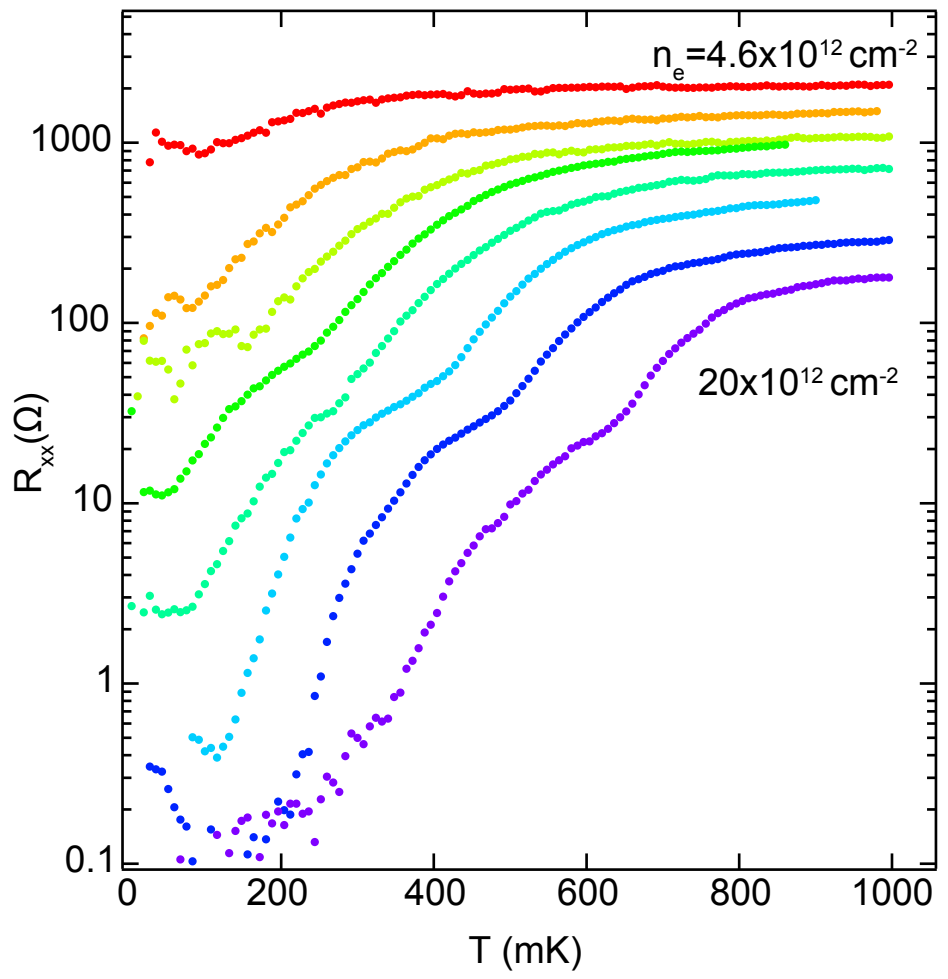

Fig S3. A plateau feature at a few 10's of Ohms is visible in this temperature dependence data, measured across the pair of voltage probes adjacent to those used for data in the main text. Curves represent longitudinal resistance $R_{xx}$ on log scale vs temperature $T$ at a series of positive-gate doping levels $n_e$ [20, 12, 8.5, 6.7, 6.1, 5.6, 5, 4.6 $\times 10^{12}$ cm$^{-2}$].

S4. Anomalous metal behavior

Figures 2B and 2D in the primary text demonstrate that, for small perpendicular magnetic fields, the resistance was nearly independent of temperature below 100 or 200 mK—the so-called anomalous metal phase(*3*). Similar effects were observed at zero perpendicular field, for densities slightly above $n_{crit}$, as seen in Fig. 2A. The lack of temperature dependence at the lowest temperatures, for densities near $n_{crit}$, can be more clearly seen when replotted vs $1/T$ (Fig S4A, analogous to Fig. 2D). For different densities, the temperature below which the resistance saturates varies from 150 mK ($5.6\times10^{12}$ cm$^{-2}$) to 50 mK ($6.1\times10^{12}$ cm$^{-2}$), although no consistent variation with density is observed.

Because the resistance saturation occurs at such low temperature, one might be tempted to ascribe the saturation to a failure of the electronic system to cool below some elevated temperature. A strong counterargument can be made based on the data in Fig. S4B, showing a dramatic difference in temperature dependence for data with nearly identical normal resistance and taken under very similar conditions.

The data in Fig. S4B highlights another observation that was not explored further in this experiment but will be subject of future investigation: The resistance saturation appeared to depend not only on the effective density, but also on the relative setting of top and bottom gate voltages (that is, on $D_\perp$). For this particular set of curves, the data corresponding to lower contact resistance (red curve) had a higher low-T resistance despite having a slightly lower resistance at 1 K, compared with the data with higher contact resistance (light blue curve).

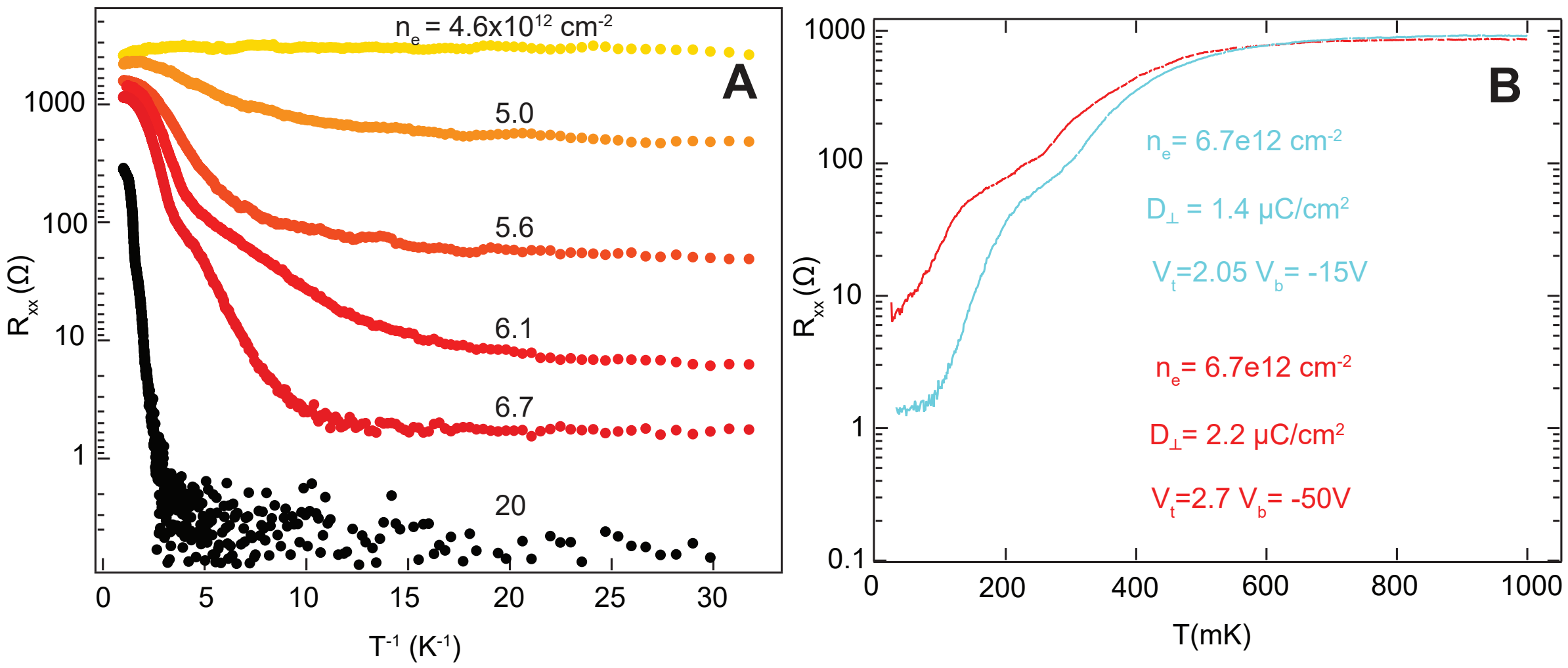

Fig S4. **A**. Data from Fig. 2A (main text) replotted vs inverse temperature to highlight behavior below 200 mK. **B.** Resistance as a function of temperature for two different displacement fields with the same carrier density, demonstrating that the electronic temperature continues to decrease with mixing chamber temperature even below 50 mK.

## S5. Nonlinear $I - V$ characteristics in the superconducting region

One common approach to characterizing superconducting systems is the measurement of nonlinear I-V characteristics, that is, the measurement of $dV/dI$ for elevated bias currents. The curves obtained from measurements of this type (see e.g. Fig. S5) had sharp peaks reminiscent of critical current measurements in a more conventional superconductor. Unfortunately, developing a clear interpretation of data like this in the present samples was not possible due to the relatively high contact resistances, low critical temperatures, and small sample sizes. Together, these factors made it too difficult to distinguish heating effects due to elevated bias currents from critical phenomena unrelated to elevated temperatures, including the various features often associated with mesoscopic superconducting samples such as multiple Andreev reflection.

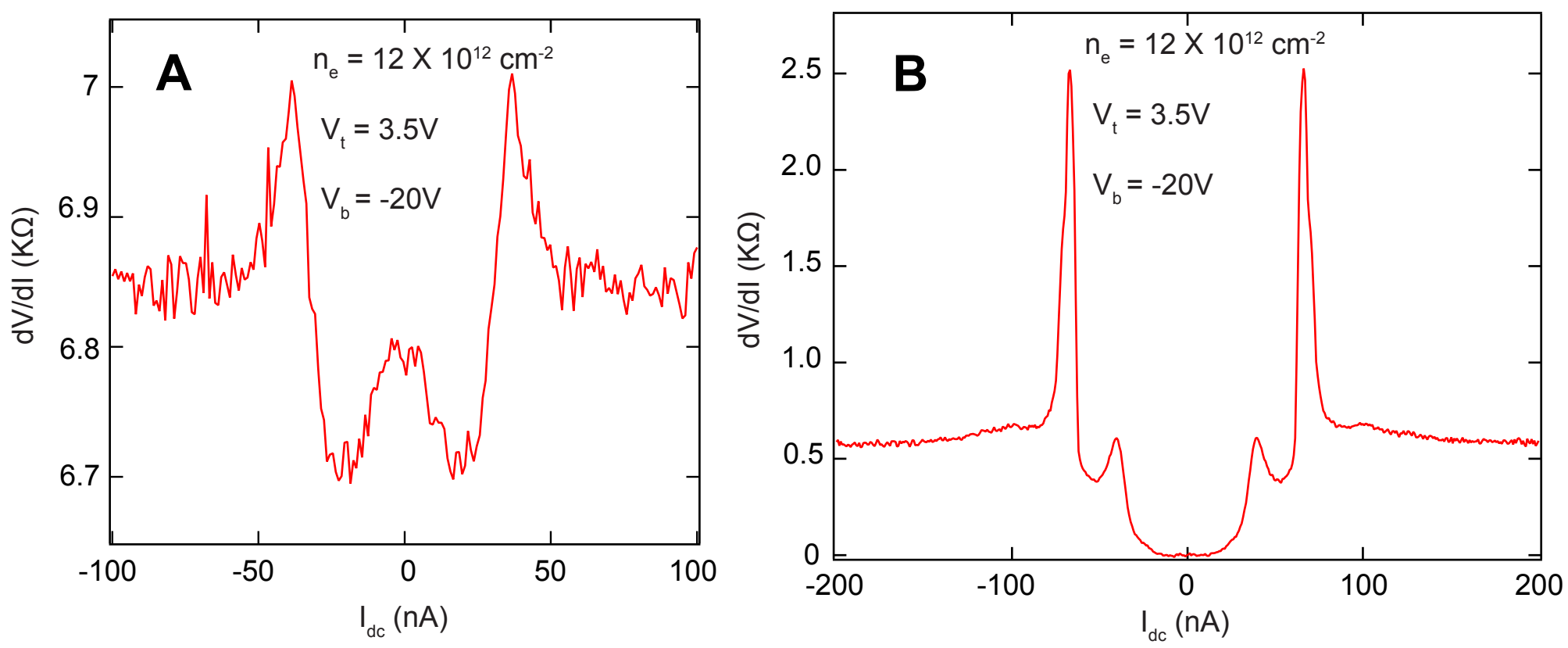

Fig S5. Two (**A**) and four (**B**) terminal differential resistance measurements for sample M1 in the superconducting regime, for the set of contacts investigated in the main text. The measurement was made with a small AC current bias (2 nA) on top of the DC current bias on the horizontal axis, using a lockin at the AC frequency to measure $dV/dI$.

## S6. Method to extract critical exponent $\alpha$ in Fig. 3

The transition from superconducting to insulating temperature dependence, as a function of $n_e$, passes through a point ($n_e = n_{crit}$) with essentially no temperature dependence, see for example Figs. 1B and 3 in the main text. To explore the possibility that this is a quantum critical point, a scaling analysis is presented in the inset to Fig. 3. That scaling analysis is described in more detail below.

The goal of this analysis is to find an exponent $\alpha$ such that multiple data sets $R_{xx}(n_e)$ taken at different temperatures collapse onto a pair of two curves (one insulating, the other metallic) when the horizontal axis is rescaled from $n_e$ to $|n_e - n_{crit}| \cdot T^{-\alpha}$. We take the following approach, illustrated in Fig. S6.

1. For a given $\alpha$, data from multiple temperatures between 100 mK and 1 K is replotted with the rescaled horizontal axis described above. Figures S6A-C show a set of data rescaled for three different $\alpha$, as an example. In order to effectively evaluate data that varies over orders of magnitude in resistance, we consider not $R_{xx}$ itself but $\ln(R_{xx})$.
2. For each rescaling of the data, best fit lines are obtained to the insulating and metallic branches (red lines in Figs. S6A-C).
3. The integrated error between the rescaled data (more precisely, the natural log of the data) and the best fit lines is then calculated, giving a quantity we refer to as "collapse error" for each $\alpha$: collapse error $= \sum_{i=0}^{n}(a + bX_i - Y_i)^2$ where a and b are intercept and slope of the best fit line respectively, $Y_i$ is the natural log of the resistance for point i and $X_i$ is the rescaled x-value for point i. The collapse errors in the insulating and superconducting branches are then added up, to get the total collapse error for a specific value of the $\alpha$.
4. This procedure is repeated for a range of $\alpha$, and a plot of collapse error versus $\alpha$ is obtained (Fig S6D). From this plot, we find that the minimum collapse error occurs for $\alpha \sim 0.8 \pm 0.1$.

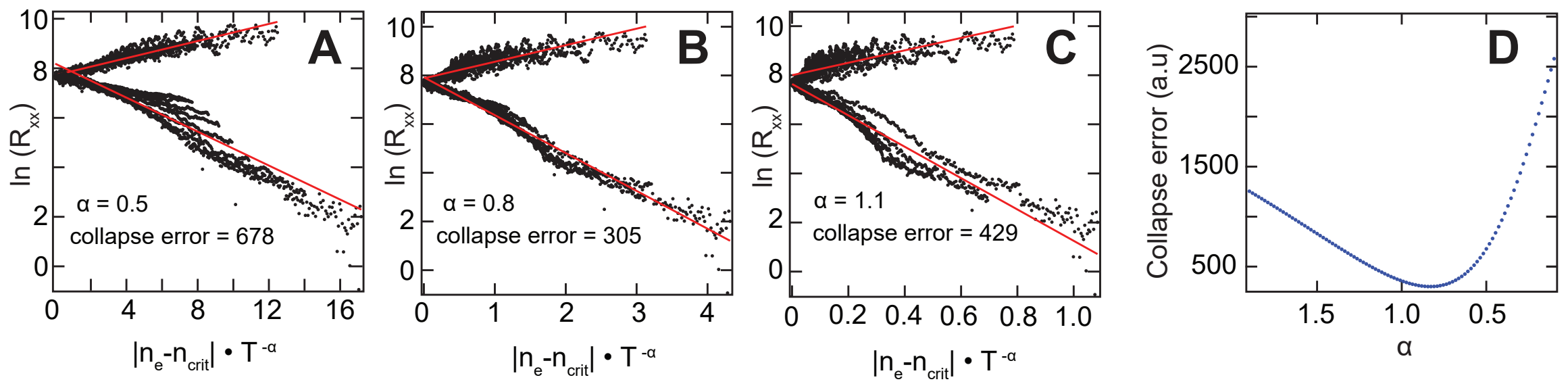

Fig S6. Explanation of the procedure used to determine an optimal scaling exponent. **A, B, C:** $R_{xx}(n_e)$ data for temperatures from 100 mK to 1 K, plotted with the rescaled horizontal axis indicated, for three values of the scaling exponent $\alpha$. Panel **B** represents the best-fit exponent, also shown in the main text (Fig. 3 inset). **D.** Collapse error quantifies the failure of the multiple-temperature datasets to collapse onto metallic and insulating branches, as described above. The minimum collapse error (best scaling) is found for $\alpha \sim 0.8 \pm 0.1$, where the error bar is determined qualitatively from the rounding of the collapse error dependence.

## S7. The edge gap and in-plane field dependence of the gap for device M1

A peculiar characteristic of all devices made by our team, following the procedure described in supplementary section S1, is the existence of a small energy gap that blocks edge transport. This gap was typically two orders of magnitude smaller than the bulk energy gap in the 2D TI state. As a result, thermally-activated edge conduction was measurable over a broad range of temperatures in the 2D TI state. Edge transport in M1 was visible from the 1 K scale, below which edge conduction was blocked, up to around 100 K, above which bulk transport was allowed, see Ref. 9 from the main text (M1 was the same device investigated in Ref. 9). Ref. 9 also clearly demonstrated the suppression of edge conduction with magnetic field.

The fact that edge conduction in M1 was only visible above 1 K, whereas superconductivity turns on only below 1 K, made it difficult to investigate the coexistence of superconductivity with 2D TI character in this device. Luckily, the edge gap was significantly smaller in M2, so some edge conduction remained even down to base temperature (20 mK) and it was relatively strong at 200 mK.

The edge gap can be characterized by non-linear conductance measurements at low temperature across the 2D TI state: the conductance is nearly zero at low bias, then at the edge of the gap the conductance rises sharply to a level that is roughly bias-independent for higher bias. The inset to Fig. 4 in the main text shows the edge gap for M2 to be around 100 μeV, not varying significantly with gate voltage across the 2D TI state. For comparison, Fig. S7A presents the analogous measurement for M1, showing an edge gap around 600 μeV with mesoscopic fluctuations but again no overall dependence on $n_e$ across the gapped region. The collapse of edge conductance as time reversal symmetry is broken by magnetic field, observed in Ref. 9 at higher temperatures, apparently is associated with a near-linear increase in the edge gap with field. The data in Fig. S7B present the growth of the edge gap in M1 with in-plane field up to 3 T, together with line cuts through the colorscale data at 0 T and 2 T.

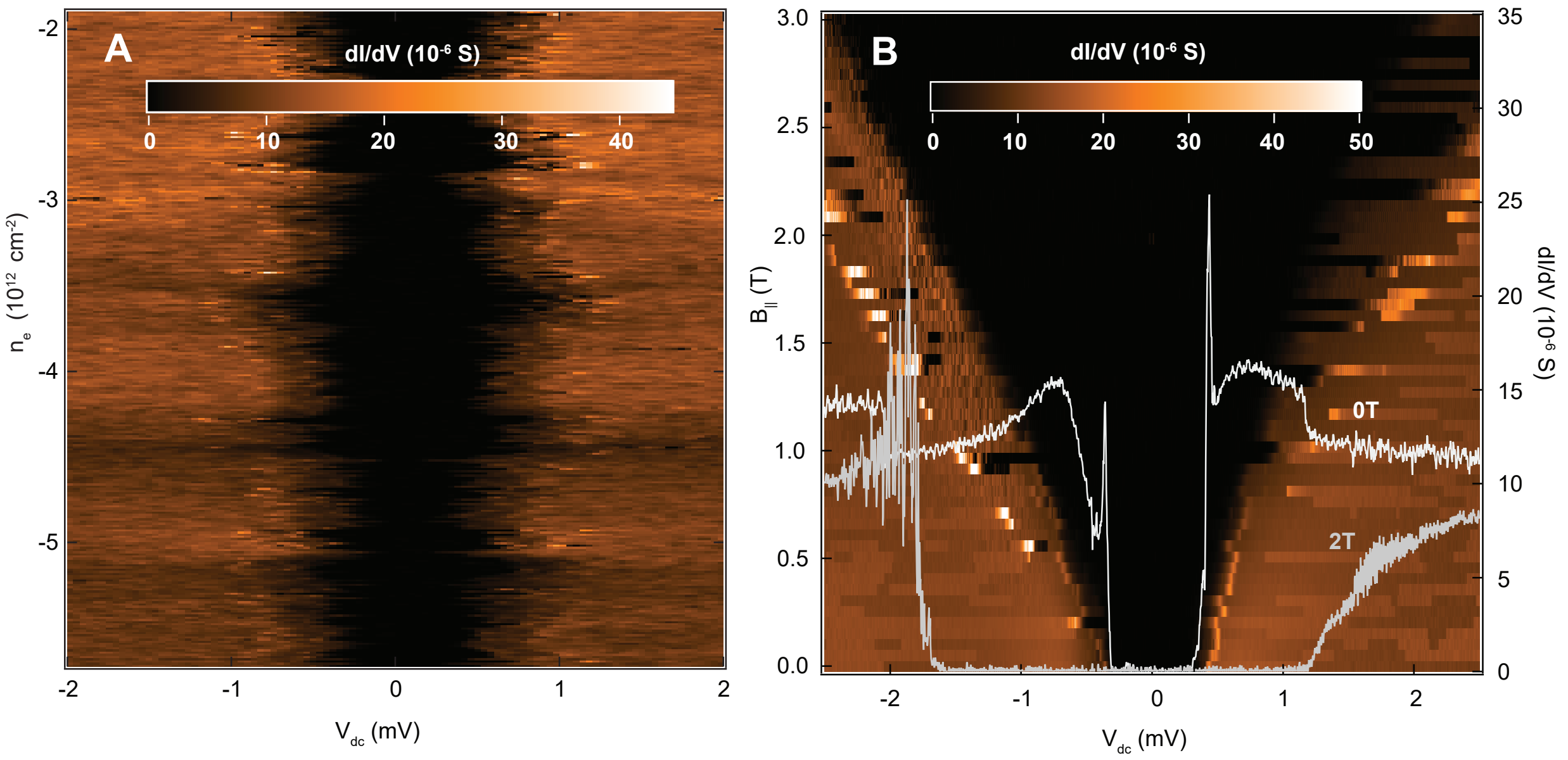

Fig S7. **A.** 2D colorscale plot of differential conductance as a function of DC bias voltage and doping level, showing a 600 μeV gap in this density range for M1. **B.** In-plane magnetic field dependence of the edge gap in M1, taken for a different setting of gate voltages where the zero-field gap was somewhat smaller than in panel A. ($T$ = 20 mK for both panels.)

## S8. Calculation of coherence length, and the criterion for identifying critical points.

The thrust of this paper is the phenomenological observation that superconductivity can be induced by mild electrostatic gating in a material that is also a quantum spin Hall insulator. We do not aim for a precise characterization of the superconductivity itself: the small size of our samples, and limited set of contact arrangements, make it difficult to extract detailed parameters (such as critical fields, temperatures, and coherence lengths) with the level of accuracy that is possible in bulk systems. Nevertheless, estimates can be made for each of these parameters; our procedure for doing so is laid out below.

A first challenge is to determine the analogue of critical fields, or temperatures, in samples where the resistance, $R_{xx}$, changes gradually as superconductivity emerges or is suppressed in the sample. To do so, it is necessary to define a particular fraction of the normal state resistance, $R_{xx}/R_N = 0.X$, that delineates the transition to superconductivity, but the choice of this fraction is somewhat arbitrary. We refer to parameters extracted through this fraction as characteristic parameters, rather than critical parameters, in order to highlight the gradual transition and any impact that might have on further analysis. Figures in the main text show parameters extracted using the characteristic fraction $0.X = 0.5$, in line with a common convention for 2D materials. Figure S8 compares the temperature dependence of characteristic out-of-plane magnetic fields, $B_{0.X}^{\perp}(T)$, extracted using fractions $0.X$ ranging from 0.1 through 0.9.

Superconducting coherence lengths are then extracted from the data in Fig. S8, for various fractions, following the two approaches mentioned in the main text. In particular, we can use either high or low temperature limits of $B_{0.X}^{\perp}(T)$ to extract a "measured" coherence length, $\xi_{meas}$. Because the coherence lengths extracted by these two approaches are very similar to each other, well within the error bar of each measurement (which itself derives primarily from the choice of characteristic fraction), we refer to a single value of coherence length simply as $\xi_{meas}$ in the main text.

The higher temperature approach to determining coherence length is based on the linear dependence of $B_{0.X}^{\perp}(T)$ data near the critical temperature $T_c$ in Figs. 2B or S8, which is consistent with the Ginzburg-Landau (GL) model for thin films that is typically used to analyze $B_{c2}$ data for 2D superconductors. In this model, the functional form of $B_{c2}^{\perp}(T)$ near $T_c$ is given by:

$$B_{c2}^{\perp} = \frac{\Phi_0}{2\pi\xi_{GL}(0)^2}\left(1 - \frac{T}{T_c}\right) ,$$

where $\Phi_0$ is the magnetic flux quantum and $\xi_{GL}(0)$ is the extrapolation of the GL coherence length to zero temperature. The slope of $B_{0.X}^{\perp}(T)$ near $T_c$ then gives $\xi_{slope} = \sqrt{\frac{-\Phi_0}{2\pi T_c \left.\frac{dB_{0.X}^{\perp}}{dT}\right|_{T_c}}}$. Without counting on the precise applicability of the GL expression over the full temperature range, it is also straightforward to estimate a value for $\xi$ in the low temperature limit, from the extrapolation of $B_{0.X}^{\perp}(T)$ to zero temperature, giving $\xi_{Bc0} = \sqrt{\frac{\Phi_0}{2\pi B_{0.X}^{\perp}(T\to 0)}}$.

The tables above Fig. S8 show that coherence lengths extracted by these two approaches are very similar to each other, and moreover that the range 0.1 to 0.9 for the critical fraction only results in a variation of $\pm 30\%$, leading to the value quoted in the main text: $\xi_{meas} = 100 \pm 30$ nm. As well, the coherence length, $\xi_d$, that might be expected in the dirty limit based purely on the critical temperature and diffusion constant is very similar to both $\xi_{slope}$ and $\xi_{Bc0}$.

| | $T_{0.X}$ (K) | $\xi_{slope}$ (nm) | $\xi_{Bc0}$ (nm) | $\xi_d$ (nm) |
|---|---|---|---|---|
| ● $B_{0.9}$ | 0.96 | 60 | 60 | 60 |
| ● $B_{0.8}$ | 0.82 | 75 | 90 | 65 |
| ● $B_{0.5}$ | 0.67 | 94 | 100 | 72 |
| ● $B_{0.3}$ | 0.6 | 103 | 105 | 76 |
| ● $B_{0.1}$ | 0.52 | 124 | 120 | 82 |

| | $T_{0.X}$ (K) | $\xi_{slope}$ (nm) | $\xi_{Bc0}$ (nm) | $\xi_d$ (nm) |
|---|---|---|---|---|
| ● $B_{0.9}$ | 1.0 | 71 | 92 | 72 |
| ● $B_{0.8}$ | 0.86 | 78 | 100 | 78 |
| ● $B_{0.5}$ | 0.71 | 95 | 115 | 86 |
| ● $B_{0.3}$ | 0.67 | 114 | 122 | 88 |
| ● $B_{0.1}$ | 0.58 | 138 | 135 | 95 |

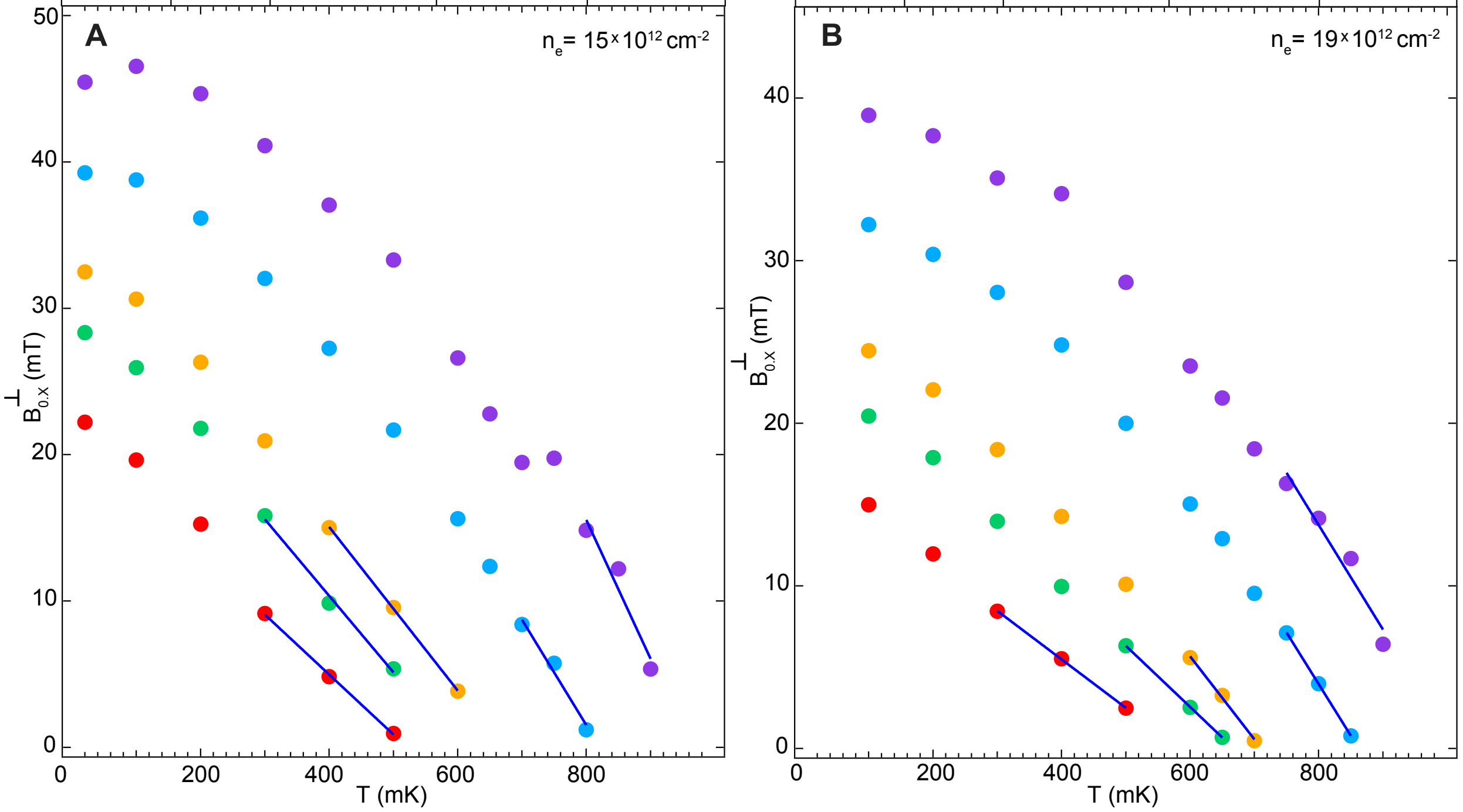

Fig S8. Temperature dependence of the characteristic field, $B^{\perp}_{0.X}(T)$, comparing characteristic fields defined using different fractions of the normal-state resistance [0.1,0.3,0.5,0.8,0.9]. Two different densities are shown (**A.** $n_e = 15\times10^{12}$ cm$^{-2}$ and **B.** $n_e = 19\times10^{12}$ cm$^{-2}$). From this temperature dependence, characteristic temperatures in the zero field limit and characteristic fields in the zero temperature limit can be estimated by extrapolation, as can the slope $\left.\frac{dB^{\perp}_{0.X}}{dT}\right|_{T_c}$. The table of values above the graphs indicates coherence lengths extracted from these data, via the two approaches described above. Also shown is the gap-based coherence length in the dirty limit, obtained from the diffusion constant and characteristic temperature as described in the main text.

## S9. Parallel magnetic field influence on critical temperature: spin-orbit analysis

Two mechanisms might explain an enhanced critical in-plane field in this material, beyond the naïve Pauli limit: a g-factor less than 2 in the expression $B_P = 1.76k_BT_c/g^{1/2}\mu_B$, or strong spin-orbit scattering. In this section, we consider the effect of fast spin-orbit scattering, and derive a spin-orbit scattering rate for the scenario where it is the dominant mechanism leading to the Pauli limit violation.

The effect of various pair-breaking mechanisms on critical temperature can be encapsulated by an implicit equation for $T_c(B_\parallel)$ that depends on a pair-breaking energy $\alpha$ (*4-6*):

$$\ln\frac{T_c(B_\parallel)}{T_{c0}} = \psi\left(\frac{1}{2}\right) - \psi\left(\frac{1}{2} + \frac{\alpha(B_\parallel)}{2\pi k_B T_c(B_\parallel)}\right)$$

where $k_B$ is the Boltzmann constant, $T_{c0}$ is $T_c$ at zero magnetic field and $\psi$ is the digamma function. Considering in-plane magnetic fields in atomically thin films (that is, with minimal orbital contribution), $\alpha$ takes the form: $\alpha \approx (g\mu_B B_\parallel)^2\tau_{so}/2\hbar$ in the case of the strong spin-orbit scattering characterized by a time $\tau_{so}$ (*4,5*).

The data in Fig. 2C inset, $T_{1/2}$ (effectively $T_c$) plotted against $B_\parallel$, are fit to the implicit function above. Since we are considering the high spin-orbit scattering limit here, the fit parameter is $\tau_{so}$ and $g$ is assumed to be 2. The fit is carried out by minimizing the function $\mathrm{F}(B_\parallel, T_c) = \ln\frac{T_c}{T_{c0}} - \psi\left(\frac{1}{2}\right) + \psi\left(\frac{1}{2} + \frac{\alpha(B_\parallel)}{2\pi k_B T_c}\right)$ evaluated for data pairs $(B_\parallel, T_{1/2})$, with respect to $\tau_{so}$. The best-fit line is shown in Fig. 2C inset, yielding an estimate for $\tau_{so}$ of around 500 fs.

Analogous to the discussion in section S8, the determination of a characteristic temperature for each $B_\parallel$ depends on the choice of fraction $R_{xx}/R_N = 0.X$. Fig. S9 compares $T_{0.X}(B_{||})$ for various $0.X$ (0.1 through 0.9, as in S8), as raw data and after scaling by characteristic fields and temperatures. Clearly, the qualitative characteristics of the data are independent of precise fraction, including the ratio by which the Pauli limit is violated, and the extracted spin-orbit times vary by only $\pm 5\%$.

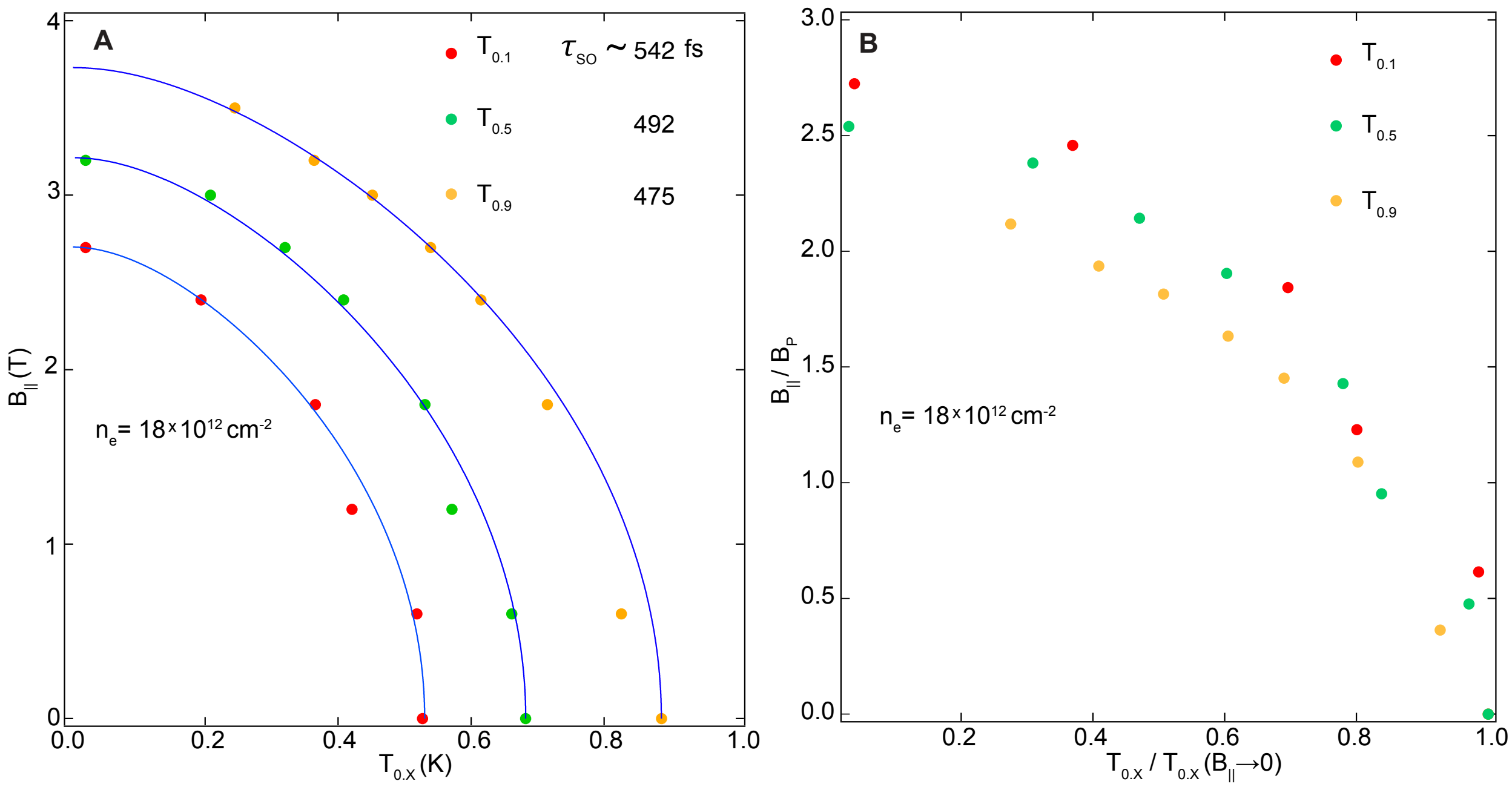

Fig S9. Effect of in-plane field $B_{||}$ on the characteristic temperature ($T_{0.X}$) extracted using different fractions 0.X [0.1,0.5,0.9]. **A**. Solid lines are the fit to the pair breaking equation in the limit of the strong spin-orbit scattering. Legend shows the extracted spin-orbit scattering time from the fitting. **B.** The same data as in **A**, with vertical and horizontal axes normalized by the Pauli limiting field and characteristic temperature at zero field respectively. The near-collapse of data for all characteristic fractions onto a single curve indicates that the choice of fraction does not qualitatively affect the interpretation.

## References

1. P. J. Zomer, M. H. D. Guimaraes, J. C. Brant, N. Tombros, B. J. van Wees, Fast pick up technique for high quality heterostructures of bilayer graphene and hexagonal boron nitride. *Applied Physics Letters* **105**, 013101 (2014).

2. S. Tang *et al.*, Quantum spin Hall state in monolayer 1T '-$WTe_2$. *Nature Physics* **13**, 683-687 (2017).
3. A. Kapitulnik, S. A. Kivelson, B. Spivak, Anomalous metals - failed superconductors. *ArXiv e-prints* **1712**, arXiv:1712.07215 (2017).
4. R. A. Klemm, A. Luther, M. R. Beasley, Theory of the upper critical field in layered superconductors. *Physical Review B* **12**, 877-891 (1975).
5. M. Tinkham, *Introduction to superconductivity*. International series in pure and applied physics (McGraw Hill, New York, ed. 2nd, 1996), pp. xxi, 454 pages.
6. K. Maki, Effect of Pauli Paramagnetism on Magnetic Properties of High-Field Superconductors. *Physical Review* **148**, 362-369 (1966).